\documentclass[floatfix, twocolumn, prl]{revtex4}

\usepackage{color}
\usepackage{multirow}
\usepackage{epsfig}
\usepackage{graphicx}
\usepackage{amsmath}
\usepackage{amssymb}
\usepackage{bm}
\usepackage{dcolumn}
\usepackage{braket}

\begin{document}

\title{Electrically tunable quantum anomalous Hall effect in $5d$ transition-metal 
adatoms on graphene}

\author{Hongbin~Zhang$^1$}
\email[corresp.\ author: ]{h.zhang@fz-juelich.de}
\author{Cesar Lazo$^2$}
\author{Stefan~Bl\"ugel$^1$}
\author{Stefan Heinze$^2$}
\author{Yuriy~Mokrousov$^1$}
\affiliation{$^1$Peter Gr\"unberg Institut and Institute for Advanced Simulation, 
Forschungszentrum J\"ulich and JARA, D-52425 J\"ulich, Germany}
\affiliation{$^2$Institute of Theoretical Physics and Astrophysics, University of Kiel, D-24098 Kiel, Germany}

\date{today}

\begin{abstract} 
The combination of the unique properties of graphene with spin
polarization and magnetism for the design of new spintronic concepts and
devices has been hampered by the small Coulomb interaction and the tiny
spin-orbit coupling of carbon in pristine graphene.  Such device
concepts would take advantage of the control of the spin degree of
freedom utilizing the widely available electric fields in electronics or
of topological transport mechanisms such as the  conjectured quantum
anomalous Hall effect.  Here we show, using first-principles methods,
that $5d$ transition-metal (TM) adatoms deposited on graphene display
remarkable magnetic properties. All considered TM adatoms possess
significant spin moments with colossal magnetocrystalline anisotropy
energies as large as 50 meV per TM atom. We reveal that the
magneto-electric response of deposited TM atoms is extremely strong and
in some cases offers even the possibility to switch the spontaneous
magnetization direction by a moderate external electric field. We
predict that an electrically tunable quantum anomalous Hall effect can 
be observed in this type of hybrid materials.

\end{abstract}

\maketitle

Spin-orbit interaction, which couples the spin degree of freedom of electrons to their 
orbital motion in the lattice, leads to many prominent physical phenomena, such as the
antisymmetric exchange interaction~\cite{bode:2007}, the colossal magnetic 
anisotropy~\cite{tosatti:2007}, or the anomalous Hall effect~\cite{Nagaosa:2010} 
in conventional ferromagnets. It is also the key interaction in the newly-found quantum 
topological phase in topological insulators, where the quantum spin Hall effect was 
observed experimentally~\cite{Konig:2007}, and the quantum anomalous Hall effect (QAHE) 
was predicted to exist~\cite{Liu:2008, Yu:2010}. Since the orbital motion can be 
manipulated with external electric fields, spin-orbit coupling (SOC) opens a route 
to dissipationless transport and to electrical control of magnetic 
properties~\cite{Weisheit:2007, Duan:2008, Rondinelli:2008}, 
playing a crucial role in future spintronics applications.

Owing to its strong spin-orbit coupling, heavy $4d$ and $5d$ TMs display fascinating
physical properties for desirable spintronic applications, especially when combined with 
non-vanishing magnetization. However, magnetism of $5d$ TMs proved 
difficult to achieve due to their more delocalized valence $d$ wavefunctions and the 
smaller intra-atomic exchange integrals compared to $3d$ TMs such as Fe, Co, or Ni. 
Low-dimensional structures, such as monolayers~\cite{Blugel:1992},
atomic wires~\cite{Mokrousov:2006}, and clusters~\cite{Piotrowski:2010}, were proposed to
facilitate the stability of $5d$  magnetism by reducing the $d$-$d$ hybridization.
Nevertheless, when deposited on substrates of noble or late transition metals, such as 
Cu, Ag, Au, and Pt, interdiffusion at interfaces is inevitable, and strong hybridization 
between the $d$ states of the substrate and of the adatoms is also destructive 
for $5d$ magnetism. From this point of view, using $sp$ substrates is more promising. 
In fact, $4d$ ferromagnetism was first observed in a Ru monolayer deposited on the 
graphite (0001) surface~\cite{Pfandzelter:1995}, which is close in its chemical and 
physical properties to graphene.

Since the day graphene was isolated and produced as a two-dimensional 
material~\cite{Novoselov:2004}, it abruptly altered the research direction of material 
science~\cite{Neto:2009} with the aim of exploring its fascinating transport properties. 
In a sense, graphene serves as a prototype of topological insulators~\cite{Hasan:2010}. 
For instance, the Berry phase of $\pi$ of electronic states in graphene induces a 
half-integer quantum Hall effect~\cite{Novoselov:2005, Zhang:2005}, while the 
existence of the quantum spin Hall effect was first suggested for pure graphene 
when SOC is taken into account~\cite{Kane:2005}. One common feature for those 
transport properties is the non-trivial topological origin, resulting in dissipationless 
charge or spin current carried by edge states with conductivity quantized in units 
of $e^2/h$. However, from the application point of view, a large external magnetic field 
is required to obtain the quantum Hall effect, and the spin-degeneracy in the quantum 
spin Hall effect makes it hard to manipulate the spin degree of freedom by controlling 
external fields. To avoid such constraints while keeping the benefit of topologically 
protected quantized topological transport, the long-sought QAHE is a perfect solution. 
The essence of the QAHE lies in the quantization of the transverse charge conductivity 
in a material with intrinsic non-vanishing magnetization. The fact that the magnetization 
in ferromagnets can be much easier handled experimentally than large magnetic fields 
makes the QAHE extremely attractive for applications in spintronics and quantum 
information. However, at present the QAHE is merely a generic theoretical concept 
for magnetically doped topological insulators~\cite{Liu:2008, Yu:2010}. Recently, Qiao 
{\it et al.}~suggested that the QAHE could also occur at comparatively low temperatures 
in graphene decorated with Fe adatoms~\cite{Qiao:2010}.


Here, we demonstrate based on first-principles theory that $5d$ TMs deposited on 
graphene are strongly magnetic, provide colossal magneto-crystalline anisotropy 
energies and exhibit topologically non-trivial band gaps due to very strong spin-orbit interaction. A generic representative
of this hybrid class of materials has a magneto-crystalline anisotropy energy of 10$-$30~meV
per TM and a quantum anomalous Hall gap in its electronic spectrum with the size of 20$-$80
meV. In connection with a large magneto-electric response of the deposited adatoms, 
we predict that within this class of systems an electrically tunable QAHE at room temperature
could be achieved experimentally. 

We have performed first-principles calculations of $5d$ TM (Hf, Ta, W, Re, Os, Ir, and Pt) 
adatoms on graphene in 2$\times$2, 3$\times$3, and 4$\times$4 supercell geometries corresponding to the deposition density of 4.7, 2.1 and 1.2~atoms/nm$^2$, respectively 
(see the Method part for details). Throughout this work the TM atoms are placed at the 
hollow sites of graphene. According to previous theoretical  studies~\cite{Chan:2008}, 
it is the most favorable absorption site with the exceptions of Pt~\cite{Yazyev:2010} 
and Ir~\cite{Zolyomi:2010}, for which the bridge site is more preferable. Among all 
systems studied we selected one prototype system, W on graphene, that we discuss 
in more detail. In the 4$\times$4 geometry,  the optimized distance between
W adatoms on the hollow site and the C plane is about 1.74~\AA, and the hollow site 
is about 0.14 eV (0.41 eV) per TM atom lower in energy than the bridge (top) site, in 
agreement with previous observations~\cite{Cretu:2010}. Applying an electric field, 
the relaxed positions change by at most 0.01~Bohr radii, and the hollow site 
remains the preferred adsorption site. 

Turning now to magnetism,  and looking first at TM atoms adsorbed in the 4$\times$4
supercell,  we find that most of the $5d$ TM atoms (except for Pt) are magnetic with 
sizable magnetic moments (see Fig.~1a)  ranging between 0.5 and 2$\mu_B$. This 
is consistent with the large magnetization energy calculated, defined as the total energy 
difference between the nonmagnetic and ferromagnetic state, reaching~e.g.~for Ta and W 
values as high as 0.56~eV and 0.34~eV, respectively. Surprisingly, the magnetic moments 
across the series of the rather isolated TM atoms do not follow Hund's first rule, i.e.~the
magnetic moments do not increase in steps of 1$\mu_B$ from Hf to Re, reaching a 
maximum at the center of the TM series and decreasing again from Re  to Pt. On the 
contrary we find the minimal spin moment for Re. This demonstrates the role of the 
hybridization of the TM $d$-orbitals with graphene. Increasing the deposition density 
studying a 3$\times$3 supercell, the results remain basically unaltered, indicating that the
$d$-$d$ hybridization between neighboring adatoms is week and that the physics is 
determined by the local TM-graphene bonding. This changes in the case of the 
2$\times$2 geometry, when the adatoms are in close vicinity to each other and 
the magnetism survives only for Ta, W, Re, and Ir atoms. 

Unique to the $5d$ TMs is the property that the SOC and the intra-atomic exchange 
are of the same magnitude. From this we anticipate a significant influence of SOC on 
the electronic properties, leading to strong anisotropies of the calculated quantities 
upon changing the direction of magnetization $\mathbf{M}$ relative to the graphene 
plane. 

The most impressive manifestation of SOC in the TM-graphene hybrid materials are 
the colossal  values of the magneto-crystalline anisotropy energy (MAE), defined as 
the total-energy difference between the magnetic states with spin moments aligned 
in the plane and out of the graphene plane. The MAE presents one of the most 
fundamental quantities of any magnetic system as its sign defines the easy 
magnetization axis and its magnitude gives an estimate on the stability of the 
magnetization with respect to temperature fluctuations $-$ a key issue for 
magnetic storage materials. A large MAE makes the magnetization very stable but 
also difficult to manipulate. 

The values of the MAE obtained in 3$\times$3 
and 4$\times$4 geometry, presented in Fig.~1b, lie in the range of 10 to 50~meV per TM. 
These are orders of magnitude larger than the values for $3d$ TMs, irrespective of whether 
they are arranged in bulk, thin films or multilayers. Similar colossal MAE magnitude has been
reported for the free-standing $4d$ and $5d$ TM chains~\cite{Mokrousov:2006} as well as
dimers of TMs~\cite{Xiao:2009}. Characteristic for the colossal MAE~\cite{tosatti:2007} is the
significant contribution from the variation of magnetic moments and corresponding
magnetization energies when changing the direction of $\mathbf{M}$, a contribution that 
is absent for conventional $3d$ magnets. For instance, for 4$\times$4 W on graphene, this
variation of the spin moment reaches as much as 0.1$\mu_B$ (Fig.~2a). By looking at the 
MAE values as the function of the band filling and deposition density in Fig.~2b, we find that,
in contrast to the spin moments, the MAE exhbits a much more irregular behavior as a function
of the band filling and depends much more sensitively on the adatom density. E.g.~for W the 
MAE changes sign depending on the density of adatoms, and has a small value in the 
4$\times$4 geometry. Such a situation provides a possibility to manipulate the 
magnetization direction with weak external perturbations, which can be particularly 
important for transport applications as we demonstrate below.

In search for ability to manipulate the magnetic properties of adatoms by external fields,
we show in Fig.~2  the dependence of the spin moments and the MAE of W adatoms in 
4$\times$4 geometry on the strength of the electric field $\mathcal{E}$, applied 
perpendicularly to the graphene layer (along the $z$-axis).  Remarkably, the spin moment 
of W displays a strong dependence on the field strength, especially when the magnetization
points out of plane. Qualitatively, this response can be characterized by a magnetoelectric
coefficient $\alpha$, which relates the change in the spin moment  to the strength of the 
$\mathcal{E}$-field in the zero-field situation: $\mu_0 \Delta \mu_S$(W) $=
\alpha\mathcal{E}$, where $\mu_0$ is the vacuum magnetic permeability constant.
For 4$\times$4 W on graphene and an out-of-plane magnetization, $\alpha_\perp$ 
amounts to about  $3\cdot 10^{-13}$~G$\cdot$cm$^2$/V, which is one order of 
magnitude larger than that in Fe thin films~\cite{Duan:2008}. For an in-plane 
magnetization, the magnetoelectric coefficient  $\alpha_\Vert$ is only 
$6\cdot 10^{-14}$~G$\cdot$cm$^2$/V, i.e.\  one order of magnitude smaller than 
$\alpha_\perp$, which underlines the strong anisotropy of this quantity.
We  observe similarly large variations of the magnetoelectric coupling strength
in other considered $5d$ TMs. The variation of orbital moment with varying 
$\mathcal{E}$-field is negligible for all systems.

\begin{figure}
\includegraphics[width=8.4cm]{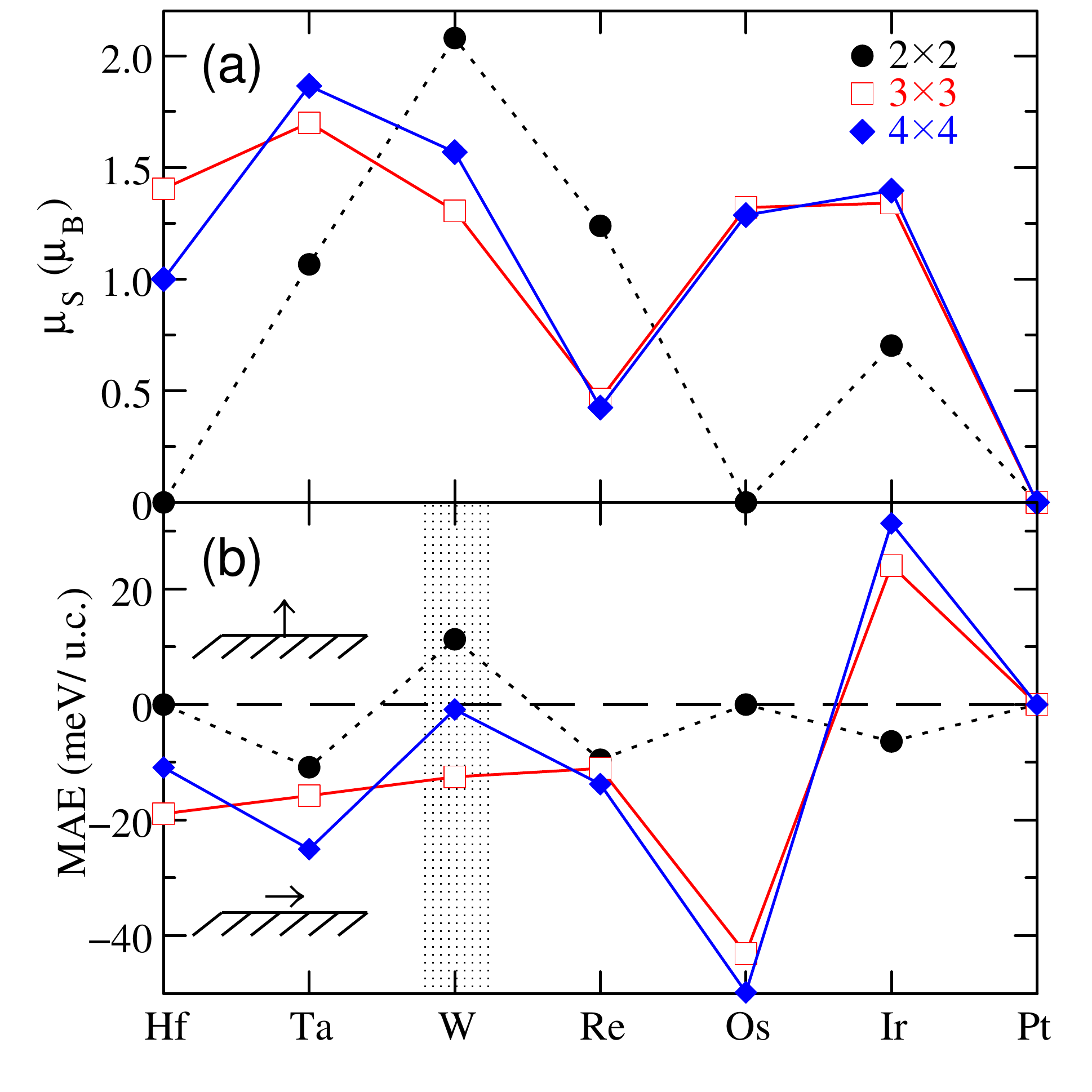}
\caption{Magnetic moments and MAE of $5d$ adatoms on graphene.
(a) Magnetic  moments due to electron spin, $\mu_S$, calculated without SOC and 
(b) the MAE of $5d$ TM 
adatoms on graphene in 2$\times$2, 3$\times$3 and 4$\times$4 superlattice geometry. 
Positive (negative)
values of MAE imply an out-of-plane (in-plane) easy axis of the spin moments, 
i.e.~perpendicular to (in) the
graphene plane.}
\end{figure}

\begin{figure}
\includegraphics[width=8.4cm]{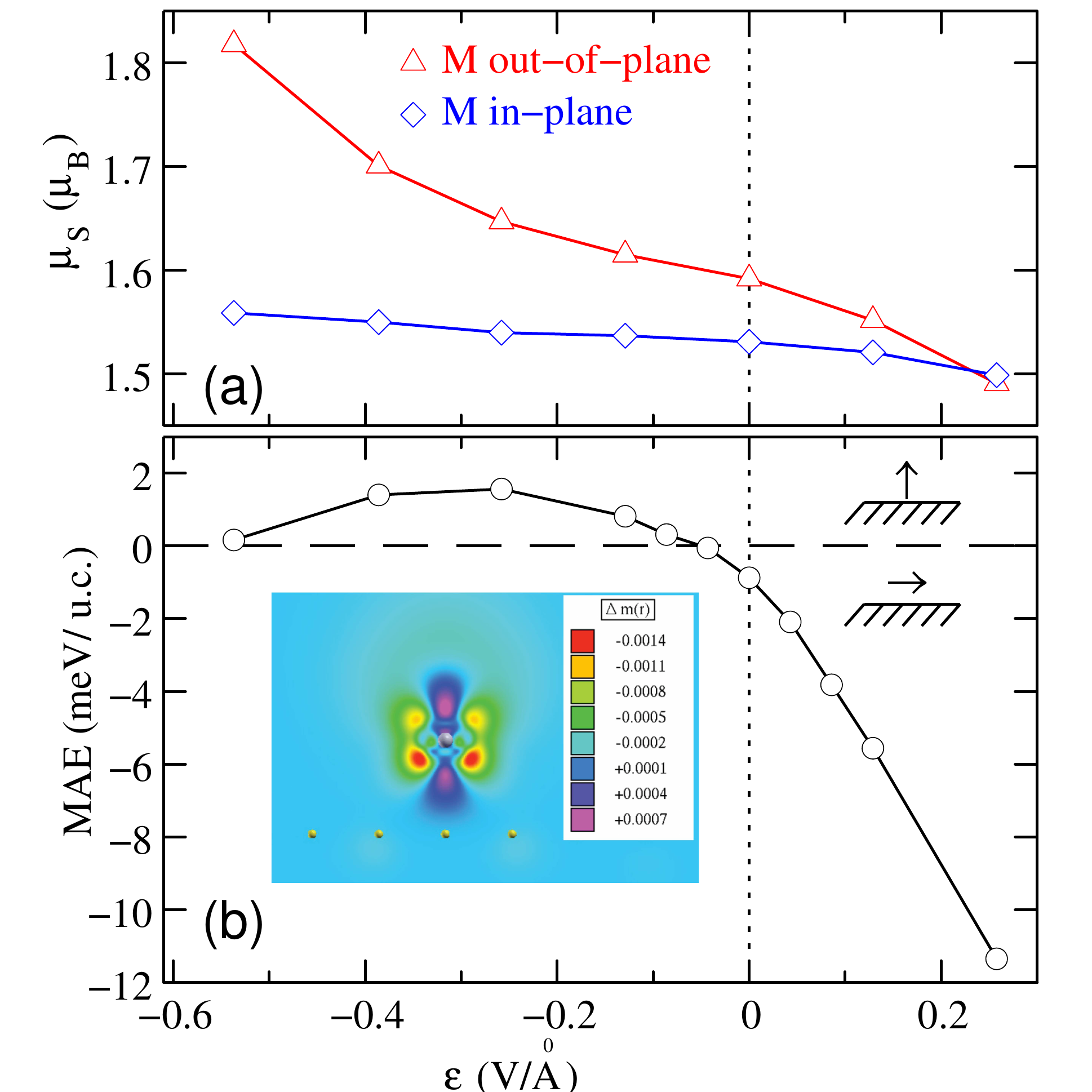}
\caption{
 Dependence of the magnetic moments (a) and magnetic anisotropy energy 
 MAE (b) of $4\times 4$ W on graphene on the strength of an external electric field.
 (a) The spin moment of W adatoms $\mu_S$ (in $\mu_B$) as a function of  the 
 strength of an external electric field $\mathcal{E}$, perpendicular to graphene surface, 
 for the out-of-plane and in-plane magnetization, respectively. 
 Negative values of $\mathcal{E}$ correspond to the 
electric field in $+z$-direction,~i.e.~the direction from graphene towards the adatoms.
(b) Dependence of the MAE on the strength of an external electric field $\mathcal{E}$, 
sign convention of MAE is consistent to Fig. 1. Inset displays the difference of spin densities 
$\Delta m(r)$ for the system without electric field and the case with a negative electric field 
of 0.13 V/\AA.}
\end{figure}

\begin{figure*}
\includegraphics[width=16cm]{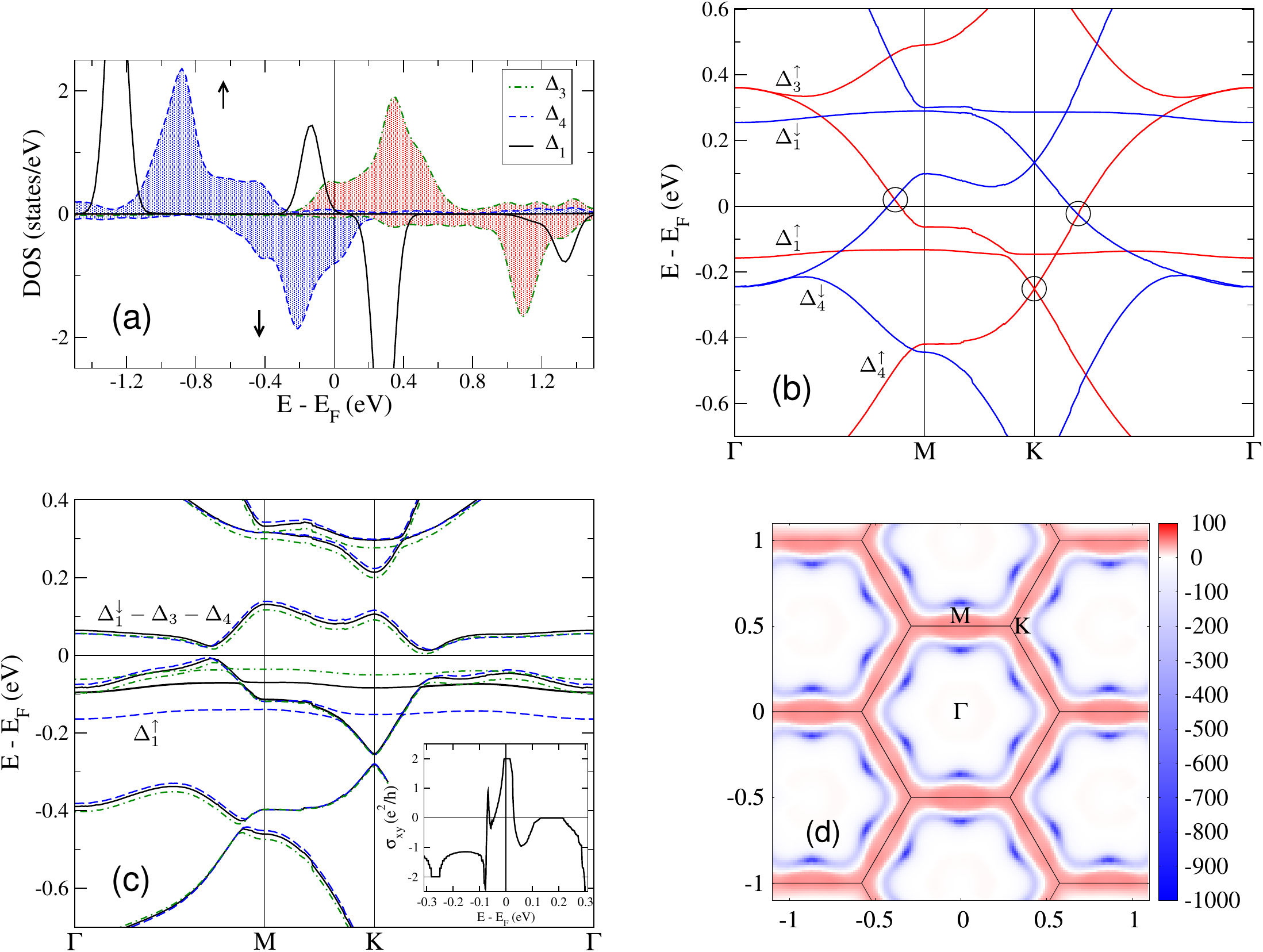}
\caption{
Electronic structure of $4\times 4$ W on graphene.
(a) Density of $\Delta_1(s,d_{z^2})$, $\Delta_3(d_{xz},d_{yz})$ and
$\Delta_4(d_{xy},d_{x^2-y^2})$ states without SOC. Up and down arrows mark spin-up 
and spin-down channels, respectively. (b) Band structure without SOC along high 
symmetry lines in the 2D Brillouin zone. Red and blue color of the bands stands for the 
spin-up and spin-down character, respectively. Circles highlight the points at which the 
gap will open when SOC is considered. (c) Band structures with SOC for $\mathbf{M}||z$: 
without electric field (solid line), with positive (dashed) and negative (dot-dashed) electric 
field of the magnitude of 0.13~V/\AA. The inset  shows the anomalous Hall conductivity 
of the system with respect to the position of the Fermi energy $E_F$ for $\mathcal{E}=0$ 
and $\mathbf{M}\Vert z$. (d) Berry curvature distribution of occupied bands in the 
momentum space (in units of $\frac{2\pi}{a}$, with $a$ as the in-plane lattice constant 
of the 4$\times$4 supercell). The Brillouin zone boundaries are marked with solid lines.}
\end{figure*}

Striking is the effect of the electric field on the magnetization direction (see Fig.~2b). 
At zero field ($\mathcal{E}=0$) the magnetization is in-plane. Applying a negative field of
magnitude of 0.05$-$0.4~V/\AA, values typical in graphene field effect transistor 
structures~\cite{Xia:2010}, the sign of the MAE is changed. This means that the 
equilibrium direction of the magnetization can be switched from in-plane ($\rm{MAE}<0$, 
for $\mathcal{E}=0$) to out-of-plane ($\rm{MAE}>0$ for $\mathcal{E}<0$). Supposing 
that the $\mathcal{E}$-field is completely screened in our system by forming a screening 
charge $\delta q$, the variation of the MAE with respect to $\delta q$, $\delta \mathrm{MAE} 
/\delta q$, reaches as much as 28 meV/e for negative electric fields, which is more 
than three times larger than that on the surface of CoPt slabs~\cite{Zhang:2009}, and 
one order of magnitude larger than that of Fe slabs~\cite{Duan:2008}. With a moderate 
out-of-plane electric field of $\pm$0.13~V/\AA~switching of the magnetization can be 
also achieved in 4$\times$4 Hf, and the MAE can be significantly altered in 4$\times$4 Os 
on graphene (by $\approx$ 10~meV) and 4$\times$4 Ir on graphene (by $\approx$ 20~meV).
This establishes $5d$ TM adatoms on graphene as a class of magnetic hybrid materials with 
high susceptibility of magnetic properties to the electric field, thus making the electrical 
control of magnetism in these systems possible.

The reason for such a strong magnetoelectric response of $5d$ TMs on graphene can 
be exemplified for the case of W in 4$\times$4 geometry. The local W $s$- and 
$d$-decomposed density of electron states without SOC, grouped into $\Delta_1(s,d_{z^2})$, 
$\Delta_3(d_{xz},d_{yz})$ and $\Delta_4(d_{xy},d_{x^2-y^2})$ contributions, is presented in
Fig.~3a. As we can see, the W spin moment of about 1.6$\mu_B$ originates from two 
occupied spin-up and two unoccupied spin-down $\Delta_1$-states, while the 
exchange-split $\Delta_3$- and $\Delta_4$-states are situated above and below $E_F$,
respectively, and almost do not contribute. Upon including SOC (e.g.~for out-of-plane spins) 
a strong hybridization between the $\Delta_1^{\downarrow}$, and $\Delta_3$ and 
$\Delta_4$ states of both spin occurs around $E_F$ (see also Fig.~3b), which results in 
a formation of hybrid bands of mixed spin and orbital character, while the 
$\Delta_1^{\uparrow}$ band remains mainly non-bonding (see Fig.~3c). When an electric 
field is applied along the $z$-axis, the states which experience most influence of the
corresponding potential change are the $\Delta_1^{\uparrow}$-states, directed 
perpendicularly to the graphene plane. Corresponding modification of the band structures 
due to the electric fields can be seen in Fig.~3c, where the $\Delta_1^{\uparrow}$ band is 
shifted downwards (upwards) by negative (positive) applied $\mathcal{E}$ fields, while the 
hybrid bands of mixed character remain almost unaffected. This causes the redistribution 
of the electrons in the $\Delta_1^{\uparrow}$-states, its hybridization with the hybrid 
band below the Fermi level and hence the variation of the magnetic moment. This can 
be visualized by the plot of the spin density difference $\Delta m(\mathbf{r})$ for the case 
with $\mathcal{E}$ = 0 and $\mathcal{E}$ = $-$0.13~V/\AA~(inset in Fig.~2b). It is obvious
that, upon applying an electric field, a certain amount of spin density is transferred from the 
$\Delta_1$ state of $d_{z^2}$ character to the $\Delta_3$ states. Owing to the difference in 
the hybridization with the graphene states, seen from the width of the corresponding peaks 
in the density of states, the $\Delta_1$ and $\Delta_3$ states have different localization 
inside the W atoms thus leading to a change in the spin moment.

At last we turn to the prediction of a stable QAHE, a manifestation of the quantization 
of the transverse anomalous Hall conductivity. It is motivated by the observation that for 
4$\times$4 W on graphene with an out-of-plane magnetization, $\Delta_3$ and 
$\Delta_4$ bands of opposite spin cross (see circles in Fig.~3b) and hybridize under 
the presence of SOC, forming a global band gap across the Brillouin zone (BZ). 
Thus, 4$\times$4 W becomes an insulator upon a spin-orbit driven metal-insulator
transition. We compute the  anomalous Hall conductivity of this system, given by 
$\sigma_{xy}=(e^2/h)\mathcal{C}$, with $\mathcal{C}$ as the Chern number of all 
occupied bands that can be obtained as a $k$-space integral
$\mathcal{C}=\frac{1}{2\pi}\int_{\rm BZ}\Omega(\mathbf{k})\,d^2k$. The integrand, 
$\Omega(\mathbf{k})$, is the so-called Berry curvature of all states below the Fermi 
level:
\begin{equation}
\Omega(\mathbf{k})=\sum_{n < E_F}\sum_{m\neq n}{\rm 2Im}
\frac{\langle\psi_{n\mathbf{k}}|v_x|\psi_{m\mathbf{k}}\rangle
   \langle\psi_{m\mathbf{k}}|v_y|\psi_{n\mathbf{k}}\rangle}
   {(\varepsilon_{m\mathbf{k}}-\varepsilon_{n\mathbf{k}})^2},
\end{equation}
where $\psi_{n\mathbf{k}}$ is the spinor Bloch wavefunction of band $n$ with corresponding
eigen energy $\varepsilon_{n\mathbf{k}}$, $v_i$ is the $i$'th Cartesian component of the 
velocity operator. The Berry curvature in reciprocal space, presented in Fig.~3d, displays 
a comparatively complex pattern, with positive contributions concentrated along the BZ
boundary, and large negative dips at certain isolated $k$-points away from the high-symmetry
points. Such non-trivial distribution of the Berry curvature, which plays a role of an effective
magnetic field in the $k$-space, is typical for complex transition-metal  
compounds~\cite{Nagaosa:2010}.

The calculated anomalous Hall conductivity as a function of the electron filling expressed 
as energy $E$ relative to the Fermi level is presented in the inset of Fig.~3c for an 
out-of-plane magnetization, $\mathbf{M}||z$, and zero $\mathcal{E}$-field. We indeed 
find that the Chern number of all occupied states acquires an integer value of $+$2. 
According to the physics of the quantum Hall effect, this will result in two dissipationless 
and topologically protected edge states on each side of a finite graphene ribbon with W 
adatoms. Remarkably, another 32~meV large QAHE gap with the $\mathcal{C}=-2$ can 
be observed at the energy of $-$0.27~eV below $E_F$, originating from the SOC-mediated
hybridization between the $\Delta_3$ and $\Delta_4$ bands of the same spin-up character,
see Figs.~3b and 3d. Such a peculiar situation suggests that the topological properties of 
the QAHE state in 4$\times$4 W  on graphene can be controlled by tuning the position of 
$E_F$~e.g.~via~deposition on an appropriate substrate. 

More importantly, at equilibrium the magnetization direction of 4$\times$4 W lies in-plane, 
rendering $\sigma_{xy}=0$ owing to the antisymmetric nature of the anomalous Hall 
conductivity with respect to the magnetization direction. From our previous observations 
it follows, however, that the direction of $\mathbf{M}$ can be conveniently switched to 
out-of-plane by applying a moderate electric field without any significant modification 
of the the electronic  structure. The  gaps remain intact (cf~Fig.~3c), its size unchanged, 
and the value of the anomalous Hall conductivity remains quantized at both energies.

Such non-trivial topological QAHE states occur also for all other $5d$ adatoms (except Pt) 
on graphene in 4$\times$4 geometry, and several of them in 2$\times$2 geometry, with 
the QAHE gaps of comparable magnitude, although positioned away from the Fermi energy. 
For example, graphene with Re deposited in 4$\times$4 geometry exhibits three 
$\mathcal{C}=+2$ QAHE gaps at $-$1, $-$0.15 and $+$0.67~eV, with the corresponding 
gap size of 38, 63 and 98~meV, respectively. Also, 4$\times$4 Ir and Os on graphene show 
two QAHE gaps ($\mathcal{C}=+2$ for Ir and $\mathcal{C}=\pm 2$ for Os) at 0.24 and 
0.47~eV for Ir, and $-$0.38 and 0.15~eV for Os, with corresponding gap size in between 
11 and 80~meV. The fact that several QAHE gaps with different Chern numbers at different
energies can occur in the same material makes these systems a rich playground for topological
transport studies. Characteristic of such systems is the leading role of the $5d$ orbitals and
strong SOC of heavy adatoms in determining the topological properties: as we can see from
Fig.~3b, the creation of the QAHE gaps can occur at the points in the Brillouin zone away from
high-symmetry points due to hybridization of bands with a strong TM character, which are
significantly altered by the TMs' SOC. This is in contrast to the cases considered 
previously~\cite{Liu:2008,Yu:2010,Qiao:2010}, in which the role of the magnetic adatoms is 
rather to insert a Rashba spin-orbit or exchange fields on the Dirac states at the Fermi energy
and high-symmetry points in the Brillouin zone. Most importantly, however, the strong SOC of
$5d$ TMs leads to an increase of the QAHE gap by an order of magnitude, when compared to 
all QAHE systems suggested before, guaranteeing a strong topological protection against 
defects, structural disorder or thermal fluctuations,  and  thus opening a route to  QAHE at
room temperature.

To summarize, we predict that graphene decorated with $5d$ transition-metals is a hybrid
material displaying remarkable magnetic and transport properties, which may be utilized
experimentally for the purposes of topological transport applications at reasonable 
temperatures and conditions. The fact that heavy transition-metals deposited on 
graphene display very stable magnetism due to colossal magneto-crystalline anisotropies 
with  exciting possibilities in engineering a robust quantum anomalous Hall effect with a 
large band gap stems from the strong spin-orbit interaction of these elements. The extreme
sensitivity of these hybrid systems to an electric field that we predicted is likely to be enhanced
when graphene is deposited on a ferroelectric substrate. Thus $5d$-graphene hybrid materials
provide a unique functionality in tailoring the required magnetic and topological properties,  
not likely to be achieved in conventional magnetic materials based on $3d$ transition-metals.
Considering the breadth of this material class including other possible heavy adatoms, such as 
the $4d$ TMs  or $4f$-elements, we anticipate that room temperature QAHE may become
possible in these systems and we encourage intense experimental research on these materials. 

\section*{Method}
The theoretical investigations are based on density functional theory~\cite{Hohenberg:1964},
which is 'first-principles' in that it requires no experimental input other than the nuclear 
charges. We apply the generalized gradient approximation~\cite{Perdew:1996} to the 
exchange-correlation potential and use the full-potential linearized augmented plane 
wave method (FLAPW) as implemented in the {\tt FLEUR} code~\cite{fleur}. The 
self-consistent calculations with SOC were carried out with the cut-off parameter 
k$_{\rm max}$ of 4.0~bohr$^{-1}$ and 144 $k$-points (for the 4$\times$4 cases) in the 
full two-dimensional Brillouin zone. The muffin-tin radius of 2.7~bohr (1.2~bohr) was 
used for TM (carbon) atoms. To calculate the transverse anomalous Hall conductivity 
accurately, we applied the Wannier interpolation technique, using the \texttt{FLEUR} 
and \texttt{Wannier90} codes to construct the maximally-localized Wannier 
functions~\cite{Souza:2002,Freimuth:2008,wannier90}. For all systems, the distance 
between the TM atoms and the rigid graphene layer was structurally optimized by total 
energy minimization neglecting SOC. We confirmed that the relaxations of the carbon 
atoms within the graphene layer are indeed very small, and neglecting them does not lead 
to any changes in the calculated quantities. TMs bond covalently with 
graphene~\cite{Chan:2008}, thus we suspect that the van der Waals 
interaction plays a minor role.

\section*{Acknowledgments}

We acknowledge discussions with Mojtaba Alaei, Frank Freimuth, Klaus Koepernik, 
Pengxiang Xu, Tobias Burnus, Gustav Bihlmayer, Marjana Le\v{z}ai\'c and Nicolae 
Atodiresei. This work was supported by the HGF-YIG Programme VH-NG-513 and 
by the DFG through Research Unit 912 and grant HE3292/7-1. Computational 
resources were provided by J\"ulich Supercomputing Centre.

\end{document}